\documentstyle[llncs]{article}

\begin{document}
\Large

\input amssym.def
\input amssym.tex     

\def\pvars{{{\cal V}_{P}}}
\def\bool{{\sf Bool}}

\newcommand{\vars}[1]{{vars(#1)}}

\def\ground{{\sf ground}}
\def\grdleq{{\sqsubseteq}}
\def\grdlub{{\sqcup}}
\def\grdglb{{\sqcap}}
\def\grddom{{\sf MSub}}
\def\grdcon{\gamma_{{\it sub}}}

\def\munify{{\it MUNIFY}} 
\def\mup{{\it MUP}} 
\def\mdown{{\it MDOWN}} 

\def\punify{{\it PUNIFY}} 
\def\pup{{\it PUP}} 
\def\pdown{{\it PDOWN}} 

\def\pgrdleq{{\Subset}}
\def\pgrdlub{{\Cup}}
\def\pgrdglb{{\Cap}}
\def\pgrddom{{\sf PSub}}
\def\pgrdcon{\Upsilon_{{\it sub}}}

\def\sub{{\sf Sub}}

\def\true{{\sl true} } 
\def\lor{\vee} 
\def\land{\wedge} 
\def\lequiv{\leftrightarrow} 
\def\implies{\rightarrow} 

% interpretation C

\def\cdom{{C}}
\def\cele{c}
\def\cbot{\bot_{C}}
\def\ctop{\top_{C}}
\def\cleq{\sqsubseteq_{C}}
\def\club{\sqcup_{C}}
\def\cglb{\sqcap_{C}}
\def\cop{{\cal C}}
\def\civ{\langle(\cdom,\cleq),\cop\rangle}
\def\cin{{\bf C}}
\def\ci{\cin=\civ}

\def\ddom{{D}}
\def\dele{d}
\def\dpele{d'}
\def\dppele{d''}
\def\dbot{\bot_{D}}
\def\dtop{\top_{D}}
\def\dleq{\sqsubseteq_{D}}
\def\dlub{\sqcup_{D}}
\def\dglb{\sqcap_{D}}
\def\dop{{\cal D}}
\def\div{\langle(\ddom,\dleq),\dop\rangle}
\def\din{{\bf D}}
\def\di{\din=\div}
\def\dabs{{\alpha}_{{D}}}
\def\dcon{{\gamma}_{{D}}}
\def\drho{{\rho}_{D}}
\def\dprho{{\rho}_{D'}}
\def\dscheme{(\civ,\dabs,\dcon,\div)}
\def\dschemen{{\sf D}}
\def\dgi{(\dcon,\cdom,\ddom,\dabs)}

\def\edom{{E}}
\def\epdom{{E'}}
\def\eele{e}
\def\epele{e'}
\def\ebot{\bot_{E}}
\def\etop{\top_{E}}
\def\eleq{\sqsubseteq_{E}}
\def\ele{\sqsubset_{E}}
\def\elub{\sqcup_{E}}
\def\eglb{\sqcap_{E}}
\def\eop{{\cal E}}
\def\eiv{\langle(\edom,\eleq),\eop\rangle}
\def\ein{{\bf E}}
\def\ei{\ein=\eiv}
\def\eabs{{\alpha}_{{E}}}
\def\econ{{\gamma}_{{E}}}
\def\erho{{\rho}_{E}}
\def\eprho{{\rho}_{E'}}
\def\escheme{(\civ,\eabs,\econ,\eiv)}
\def\eschemen{{\sf E}}
\def\egi{(\econ,\cdom,\edom,\eabs)}

\def\pdom{{P}}
\def\pele{p}
\def\pbot{\bot_{P}}
\def\ptop{\top_{P}}
\def\pleq{\sqsubseteq_{P}}
\def\plub{\sqcup_{P}}
\def\pglb{\sqcap_{P}}
\def\pop{{\cal P}}
\def\piv{\langle(\pdom,\pleq),\pop\rangle}
\def\pin{{\bf P}}
\def\pi{\pin=\piv}
\def\pabs{{\alpha}_{{P}}}
\def\pcon{{\gamma}_{{P}}}
\def\prho{{\rho}_{P}}
\def\pscheme{(\civ,\pabs,\pcon,\piv)}
\def\pschemen{{\sf P}}
\def\pgi{(\pcon,\cdom,\pdom,\pabs)}

\def\rdom{{R}}
\def\rele{r}
\def\rbot{\bot_{R}}
\def\rtop{\top_{R}}
\def\rleq{\sqsubseteq_{R}}
\def\rlub{\sqcup_{R}}
\def\rglb{\sqcap_{R}}
\def\rop{{\cal R}}
\def\riv{\langle(\rdom,\rleq),\rop\rangle}
\def\rin{{\bf R}}
\def\ri{\rin=\riv}
\def\rabs{{\alpha}_{{R}}}
\def\rcon{{\gamma}_{{R}}}
\def\rrho{{\rho}_{R}}
\def\rscheme{(\civ,\rabs,\rcon,\riv)}
\def\rschemen{{\sf R}}
\def\rgi{(\rcon,\cdom,\rdom,\rabs)}

\def\udom{{U}}
\def\uele{u}
\def\ubot{\bot_{U}}
\def\utop{\top_{U}}
\def\uleq{\sqsubseteq_{U}}
\def\ulub{\sqcup_{U}}
\def\uglb{\sqcap_{U}}

\def\fst{{\sf fst}}
\def\snd{{\sf snd}}

\def\redde{{\it red}_{\ddom\times\edom}}
\def\equivde{\equiv_{\ddom\times\edom}}

\newcommand {\fcomp}{\cdot}
\newcommand {\Vect}[1]{\mbox{$#1_1,#1_2,\cdots,#1_n$}}

\newcommand{\point}[1] {\circled{#1}}
\newcommand{\edge}[2]{{#1
      \raisebox{-.2ex}{$\leftarrow\!\!\!\!\bullet$}#2}}
\newcommand{\stackitem}[3]{\mbox{$\parallel\edge{#1}{#2},#3\parallel$}}

\def\syneq{\equiv}

\def\thus{\hbox to 12pt{.\raise 4pt \hbox{.}.}}

\def\therefore{\mbox{$\mathrel{.\dot{~}.}$~}}
\def\because{\mbox{$\mathrel{\dot{~}.\dot{~}}$~}}

\def\modes{{\sf MO}}
\def\modeg{{\sf g}}
\def\modeu{{\sf u}}
\def\mode{{\sf m}}
\def\modeleq{\unlhd}
\def\modelub{\bigtriangledown}
\def\modeglb{\bigtriangleup}
\def\modecon{\gamma_{{\it term}}} 

\def\pmodes{{\sf PM}}
\def\pmodepre{{\ll}}
\def\pmodeequiv{{\cong}}
\def\pmodeleq{\preceq}
\def\pmodelub{\oplus}
\def\pmodeglb{\otimes}
\def\pmodecon{\Upsilon_{{\it term}}}

\def\dcal{{\cal D}}
\def\scal{{\cal S}}
\def\rcal{{\cal R}}

\def\lcal{{\cal L}}
\def\vcal{{\cal V}}
\def\wcal{{\cal W}}
\def\gcal{{\cal G}}
\def\ncal{{\cal N}}
\def\ecal{{\cal E}}
\def\ucal{{\cal U}}
\def\ycal{{\cal Y}}
\def\zcal{{\cal Z}}

\def\fail{{\sf fail}}

\def\mbeta{\beta^\flat}
\def\meta{\eta^\flat}
\def\msigma{\sigma^\flat}
\def\mtheta{\theta^\flat}
\def\mzeta{\zeta^\flat}

\def\pbeta{\beta^\sharp}
\def\peta{\eta^\sharp}
\def\psigma{\sigma^\sharp}
\def\ptheta{\theta^\sharp}
\def\pzeta{\zeta^\sharp}

\def\dsem{F^\diamond_{P}}
\def\ddom{\dcal^\diamond}
\def\dgamma{\gamma^\diamond}
\def\dsigma{\sigma^\diamond}
\def\dtheta{\theta^\diamond}
\def\dbigx{X^\diamond}
\def\dbigy{Y^\diamond}
\def\dsqsubseteq{\sqsubseteq^\diamond}
\def\dsqcap{\sqcap^\diamond}
\def\dsqcup{\sqcup^\diamond}
\def\dtop{\top^\diamond}
\def\dbot{\bot^\diamond}

\def\sem{F_{P}}
\def\dom{\dcal}
\def\unify{unify}

\def\initstacks{{\scal}_{0}}
\def\finalstacks{{\scal}_{\infty}}
\def\stacks{{\scal}}

\def\edges{\ecal_{P}}
\def\points{\ncal_{P}}
\def\lfp{{\em lfp}}

\newcommand{\asub}   {\overline{ASub}}
\newcommand{\asubgamma} {\overline{\gamma}}
\newcommand{\asuborder} {\overline{\sqsubseteq}}
\newcommand{\asubbot}  {\overline{\bot}}
\newcommand{\asubtop}  {\overline{\top}}
\newcommand{\asubcup}  {\overline{\sqcup}}
\newcommand{\asubcap}  {\overline{\sqcap}}

\def\asubf {\overline{ASub}}
\def\asubgammaf{\overline{\gamma}}
\def\asubcupf{\overline{\sqcup}}
\def\asubidf{\overline{\epsilon}}
\def\munifyf{\overline{unify}}
\def\dunifyf{\widehat{unify}}

\newcommand{\pgname}[1] {{\sl #1}}
\newcommand{\setof}[1] {{#1}^{\diamond}}
\def\term{{\sf Term}}
\def\atom{{\sf Atom}}

\def\cons{{\sf Cons} } 
\def\pars{{\sf Para} } 
\def\mono{{\sf Mono} } 
\def\patt{{\sf Pat}}
\def\prop{{\sf Prop} } 

\def\tint{{\sl int}}
\def\tnat{{\sl nat}}  
\def\tlist{{\sl list}} 
\def\ttree{{\sl tree}}

\def\mtsubpreorder{\mbox{$\preceq\hspace{-0.65em}\raisebox{0.8ex}{$\scriptstyle\shortparallel$}$}}
\def\mtsubequiv{\mbox{$\approx\hspace{-0.65em}\raisebox{0.3ex}{$\scriptstyle\shortparallel$}$}}
\def\mtsubdom{{\sf MTS}}
\def\mtsubleq{\mbox{$\sqsubseteq\hspace{-0.65em}\raisebox{0.9ex}{$\scriptstyle\shortparallel$}$}}
\def\mtsublub{\mbox{$\sqcup\hspace{-0.4em}\raisebox{0.7ex}{$\scriptscriptstyle =$}$}}
\def\mtsubbiglub{\mbox{$\bigsqcup\hspace{-0.55em}\raisebox{0.7ex}{$\scriptscriptstyle =$}$}}
\def\mtsubglb{\mbox{$\sqcap\hspace{-0.4em}\raisebox{0.7ex}{$\scriptscriptstyle =$}$}}
\def\mtsubbot{\mbox{$\bot\hspace{-0.65em}\raisebox{0.7ex}{$\scriptscriptstyle =$}$}}
\def\mtsubtop{\mbox{$\top\hspace{-0.65em}\raisebox{0.7ex}{$\scriptscriptstyle =$}$}}

% start of definitions for groundness analysis

\def\gtheta{\overline{\theta}}
\def\gsigma{\overline{\sigma}}
\def\gzeta{\overline{\zeta}}
\def\geta{\overline{\eta}}
\def\gdelta{\overline{\delta}}
\def\gbot{\overline{\bot}} 
\def\gtop{\overline{\top}} 
\def\glub{\overline{\sqcup}} 
\def\gglb{\overline{\sqcap}} 
\def\gleq{\overline{\sqsubseteq}} 
\def\ggamma{\overline{\gamma}} 
\def\gsub{\overline{Sub}} 
\def\gunify{\overline{{\it UNIFY}}} 
\def\gpvars{{\sf V}_{\pg}}
\def\gbotplus{\gbot^{+}}
\def\gtopplus{\gtop^{+}}
\def\glubplus{\glub^{+}}
\def\gglbplus{\gglb^{+}}
\def\gleqplus{\gleq^{+}} 
\def\ggammaplus{\ggamma^{+}} 
\def\gsubplus{\gsub^{+}}
\def\gupplus{\overline{up}} 
\def\gdownplus{\overline{down}}
\def\gpvarsplus{\gpvars^{+}}

% end of definitions for groundness analysis

% environment definitions

\newtheorem {algorithm}{Algorithm} 

\newcommand {\definedas}{\stackrel{\em def}{=}} 
\newcommand {\Vector}[2]{{#1_1,\cdots,#1_#2}} 
\newcommand {\values}{\mapsto} 
\newcommand {\first}[1]{{\it #1}} 
\newcommand {\comments} [1]{}
\newcommand {\delete} [1]{}
\newcommand{\insertion}[1]{#1}
\newcommand{\sinsertion}[1]{#1}

% logic program 

\def\pg{{\sf P}} 
\def\syneq{\equiv} 
\def\eqs{{\sf Eqn} } 
\def\unify{{\it unify}} 
\def\lfp{{\it lfp}} 
\def\fail{{\it fail}} 
\def\mgu{{\it mgu}}
\def\func{{\Sigma} } 
\def\allvars{{\cal V} } 
\def\ucal{{\sf U} } 
\def\vcal{{\sf V} } 
\def\wcal{{\sf W} } 
\def\pred{{\Pi} } 
\def\restrict{\mbox{$\mid\hspace{-0.45em} \raise 4pt
        \hbox{$\scriptscriptstyle \setminus$}$}}

% logical connectives

\def\true{{\sl t} } 
\def\false{{\sl f}} 
\def\lor{\vee} 
\def\land{\wedge} 
\def\lequiv{\leftrightarrow} 
\def\lbigand{\bigwedge} 
\def\lbigor{\bigvee} 
\def\implies{\rightarrow} 

% general reference to abstract interpretation

% universeral p.a.i.

\def\upgamma{{\Upsilon}} 
\def\upassign{{\bf A}}
\def\uemul{{\propto}}
\def\upopb{{\cal P}^\ell}

\def\updom{{\bf P}} 
\def\upleq{\Subset} 
\def\upbot{{\bot}_{\bf P}} 
\def\uptop{{\top}_{\bf P}} 
\def\uplub{\Cup} 
\def\upglb{\Cap} 
\def\upai{{\Bbb P}} 
\def\upop{{\cal P}}
\def\upele{{\bf p}}

\def\uppreorder{{\preceq}}
\def\upequiv{{\thickapprox}}
\def\upsubsumed{{\unlhd}}

\def\upbdom{{\bf E}}
\def\upbool{{\bf B}}
\def\upbele{{\bf e}}
\def\upcc{{\bf b}}
\def\upbecc{{\bar{{\bf e}}}}
\def\bphi{{\Phi}}
\def\upsetS{{\Phi}_1}
\def\upsetR{{\Phi}_2}

% universal m.a.i.

\def\umleq{\preccurlyeq} 
\def\umlub{\curlyvee} 
\def\umglb{\curlywedge} 
\def\umbot{{\bot}_{\bf M}} 
\def\umtop{{\top}_{\bf M}} 

\def\umai{{\Bbb M}} 
\def\umop{{\cal M}}
\def\umele{{\bf m}}
\def\umgamma{{\gamma}} 
\def\umdom{{\bf M}} 

% universal c.a.i. 

\def\ucleq{\sqsubseteq} 
\def\ucbot{\bot} 
\def\uctop{\top} 
\def\uclub{\sqcup} 
\def\ucbiglub{\bigsqcup} 
\def\ucglb{\sqcap} 

\def\ucdom{{\bf C}}
\def\uci{{\Bbb C}} 
\def\ucop{{\cal C}}
\def\ucele{{\bf c}}

% framework c.a.i.

\def\cbot{{\emptyset}} 
\def\ctop{{Sub}} 
\def\club{{\cup}} 
\def\cglb{{\cap}} 
\def\cleq{{\subseteq}} 
\def\csub{{\wp(Sub)}} 
\def\cunify{{\it UNIFY}} 
\def\fci{{I}}

% framework m.a.i.

\def\fmlub{{\sqcup}} 
\def\fmleq{{\sqsubseteq}} 
\def\fmsub{{\it ASub}} 
\def\fmunify{{\it AUNIFY}} 
\def\fmgamma{{\gamma}}
%\def\fmai{{I}} 

% other things

\def\bBbbk{{\bar{\Bbbk}}}

\def\inputinfo{{\sf In}}
\def\outputinfo{{\sf Out}}

\bibliographystyle{plain}

\title{A Polymorphic Groundness Analysis of Logic Programs}
\author{Lunjin Lu}
\institute{Department of Computer Science\\
        University of Waikato\\ Hamilton, New Zealand\\
        EMail: lunjin@cs.waikato.ac.nz}
\date{}

\maketitle

\begin{abstract}
A polymorphic analysis is an analysis whose input and output contain
parameters which serve as placeholders for information that is unknown
before analysis but provided after analysis. In this paper, we present
a polymorphic groundness analysis that infers parameterised groundness 
descriptions of the variables of interest at a program point. 
The polymorphic groundness analysis is designed by replacing two primitive 
operators used in a monomorphic groundness analysis and is shown to 
be as precise as the monomorphic groundness analysis for any possible 
values for mode parameters. Experimental results of a prototype 
implementation of the polymorphic groundness analysis are given.

\end{abstract}

\section{Introduction} \label{sec:intro}
Groundness analysis is one of the most important dataflow analyses for
logic programs. It provides answers to questions such as whether, at a
program point, a variable is definitely bound to a term that contains no
variables.  This is useful not only to an optimising compiler that attempts
to speed up unification but also to other program manipulation tools that
apply dataflow analyses. Groundness analysis has also been used to improve
precisions of other dataflow analyses such as sharing
analysis~\cite{Sondergaard:86,Jacobs89:NACLP,Jacobs:JLP92,CodishDY91,CortesiF91,MuthukumarH91,SundararajanC92,King}
and type analysis~\cite{Lu95}.  There have been many methods proposed for
groundness
analysis~\cite{Mellish:85,Dart91,CodishDY91,SundararajanC92,CortesiFW_JLP96,Mar-Son:loplas93,CortesiFW_JLP96,Arm-Mar-Sch-Son:SAS94,Arm-Mar-Sch-Son:SCP98,Charlier92,LeCharlier:1993:GAP,pvh:jlp95b,Codish:ILPS93,CD95:prop,Son:FSTTCS96}.

This paper present a new groundness analysis, called polymorphic 
groundness analysis, whose input and output are parameterised 
by a number of mode parameters. These mode parameters represent 
groundness information that is not available before analysis 
but can be provided after analysis. When the groundness information is
provided, the result of the polymorphic groundness analysis can 
then be "instantiated".  Consider the following program
\begin{verbatim}
    p(X,Y) :- q(X,Y), X<Y, ...
    q(U,U).
\end{verbatim} 
and the goal $p(X,Y)$ with mode parameters $\alpha$ and $\beta$ being
respectively the groundness of $X$ and $Y$ prior to the execution of
the goal. The polymorphic groundness analysis infers that at the
program point immediately before the built-in call $X<Y$, the
groundness of both $X$ and $Y$ are the greatest lower bound of
$\alpha$ and $\beta$ implying that $X$ and $Y$ are in the intersection
of the sets of terms described by $\alpha$ and $\beta$.  The result
can be instantiated when the values of $\alpha$ and $\beta$ become
available. It can also be used to infer sufficient conditions for
safely removing run-time checks on the groundness of the operands of
the built-in predicate $"<"$.

The polymorphic groundness analysis is performed by abstract
interpretation~\cite{Cousot:Cousot:77,Cousot:Cousot:79,Cousot:JLP92}. Abstract
interpretation is a methodology for static program analysis whereby a
program analysis is viewed as the execution of the program over a
non-standard data domain. A typical analysis by abstract
interpretation is monomorphic whereby input information about the
program is not parameterised and the program has to be analysed
separately for different input information.  The polymorphic
groundness analysis is one of a particular class of program analyses
whereby input and output of a program analysis contain parameters
which express information that is unknown before analysis but may be
provided after analysis. Polymorphic program analyses have advantages
over monomorphic program analyses since more general result can be
obtained from a polymorphic analysis. Firstly, the result of a
polymorphic analysis is reusable. A sub-program or a library program
that may be used in different places need not to be analysed
separately for its different uses. Secondly, polymorphic program
analyses are amenable to program modifications since changes to the
program does not necessitate re-analyses of the sub-program so long as
the sub-program itself is not changed.

The polymorphic groundness analysis is a polymorphic abstract
interpretation that formalises a polymorphic analysis as a generalisation
of a possibly infinitely many monomorphic analyses~\cite{LuJLP98}.  It is
obtained from a monomorphic groundness analysis by replacing monomorphic
description domains with polymorphic description domains and two primitive
operators for the monomorphic groundness analysis with their polymorphic
counterparts. It is proven that for any possible assignment of values for
parameters, the instantiated results of the polymorphic groundness analysis
is as precise as that of the monomorphic groundness analysis corresponding
to the assignment.

  An abstract interpretation framework is a generic abstract semantics
that has as a parameter a domain, called an abstract domain, and a
fixed number of operators, called abstract operators, associated with
the abstract domain. A particular analysis corresponds to a particular
abstract domain and its associated abstract operators. Usually,
specialising a framework for a particular analysis involves devising
an abstract domain for descriptions of sets of substitutions, called
abstract substitutions, and corresponding abstract operators on
abstract substitutions. One important abstract operator commonly
required is abstract
unification~\cite{CodishDY91,Jones:Sondergaard:87} which mainly
abstracts the normal unification. Another important abstract operator
commonly required is an operator that computes (an approximation of)
the least upper bound of abstract substitutions. The polymorphic
groundness analysis will be presented as an abstract domain for
polymorphic groundness descriptions of sets of substitutions together
with its associated abstract unification operator and least upper
bound operator as required by the framework in~\cite{LU94}. The
adaptation to the frameworks in
~\cite{CodishDY91,Jacobs:JLP92,MarriottSJ94,Nilsson:88} needs only
minor technical adjustments since the functionalities required of the
above two abstract operators by these frameworks and that
in~\cite{LU94} are almost the same. The adaptation to the frameworks
in~\cite{Bruynooghe91,Kanamori:93,Mellish:87} needs more technical
work but should not be difficult because most functionalities of the
abstract operators in these frameworks are covered by the
functionalities of the abstract unification operator and the least
upper bound operator
in~\cite{CodishDY91,Jacobs:JLP92,LU94,Nilsson:88}.

The remainder of the paper is organised as
follows. Section~\ref{sec:pre} introduces basic notations, abstract
interpretation and an abstract interpretation framework of logic
programs based on which we present our polymorphic groundness
analysis.  Sections~\ref{sec:pai} recalls the notion of polymorphic
abstract interpretations.  Section~\ref{sec:mono} reformulates the
monomorphic groundness analysis and section~\ref{sec:poly} presents
the polymorphic groundness analysis.  Section~\ref{sec:imp} contains
results of the polymorphic groundness analysis on some example
programs and provides performance results of a prototype
implementation of the polymorphic groundness analysis.
Section~\ref{sec:conc} concludes the paper with a comparison with
related work on groundness analysis of logic programs.

\section{Preliminaries} \label{sec:pre}

This section recalls the concept of abstract interpretation and an
abstract interpretation framework based on which we will present
our polymorphic groundness analysis. 
The reader is assumed to be familiar with
terminology of logic programming~\cite{Lloyd:87}.

\subsection{Notation}

Let $\func$ be a set of \first{function symbols}, $\pred$ be a set
of \first{predicate symbols}, $\allvars$ be a denumerable set of
variables.  \sinsertion{$f/n$ denotes an arbitrary function
symbol, and capital letters denote variables.}  $\term$ denotes
the set of \first{terms} that can be constructed from $\func$ and
$\allvars$. \sinsertion{$t,t_i$ and $f(\Vector{t}{n})$ denote
arbitrary terms.} $\atom$ denotes the set of \first{atoms}
constructible from $\pred$, $\func$ and
$\allvars$. \sinsertion{$a_1$ and $a_2$ denote arbitrary atoms.}
\sinsertion{$\theta$ and $\theta_i$ denote substitutions.}  Let
$\theta$ be a \first{substitution} and
\mbox{$\vcal\subseteq\allvars$}.  $dom(\theta)$ denotes the domain
of $\theta$.  $\theta\restrict\vcal$ denotes the restriction of
$\theta$ to $\vcal$. As a convention, the function composition
operator $\circ$ binds stronger than $\restrict$. For instance,
$\theta_1\circ\theta_2\restrict\vcal$ is equal to
$(\theta_1\circ\theta_2)\restrict\vcal$.  An \first{expression}
$O$ is a term, an atom, a literal, a clause, a goal
etc. $\vars{O}$ denotes the set of variables in $O$. 
The range $range(\theta)$ of a substitution $\theta$ 
is $range(\theta)\definedas\cup_{X\in dom(\theta)}\vars{\theta(X)}$.
 
An \first{equation} is a formula of the form $l=r$ where either
$l,r\in \term$ or $l,r\in\atom$.  The set of all equations is
denoted as $\eqs$.  Let $E\in\wp(\eqs)$.  $E$ is in \first{solved
form} if, for each equation $l=r$ in $E$, $l$ is a variable that
does not occur on the right side of any equation in $E$.  For a
set of equations $E\in\wp(\eqs)$, $\mgu: \wp(\eqs)\mapsto
Sub\cup\{\fail\}$ returns either a most general unifier for $E$ if
$E$ is unifiable or $\fail$ otherwise, where $Sub$ is the set of
substitutions. $\mgu(\{l=r\})$ is sometimes written as
$\mgu(l,r)$. Let $\theta\circ\fail\definedas\fail$ and
$\fail\circ\theta\definedas\fail$ for any $\theta\in
Sub\cup\{\fail\}$.  There is a natural bijection between
substitutions and the sets of equations in solved
form. $eq(\theta)$ denotes the set of equations in solved form
corresponding to a substitution $\theta$.
$eq(\fail)\definedas\fail$.

We will use a renaming substitution $\Psi$ which renames a
variable into a variable that has not been encountered before. Let
$\pvars$ be the set of program variables of interest. $\pvars$ is
usually the set of variables in the program.

\subsection{Abstract Interpretation} 

Suppose that we have the \pgname{append} program in
figure~\ref{fig:append}.
\begin{figure}
\begin{center}
\begin{tabular}{lcll}  
$append([~],L,L)$ & & & \%C1\\ 
$append({[H|L_1]},L_2,[H|L_3])$ & $\leftarrow$ &
                $append(L_1,L_2,L_3)$ & \%C2\\ 
\end{tabular}
\end{center}
 \caption{\label{fig:append} Logic program $append$} 
\end{figure}  

The purpose of groundness analysis is to find answers to such questions as in the
following.
\begin{quotation} 
If ${L_1}$ and ${L_2}$ are ground before
${append(L_1,L_2,L_3)}$ is executed, will 
${L_3}$ be ground after ${append(L_1,L_2,L_3)}$ is successfully executed?
\end{quotation} 
Abstract interpretation performs an analysis by
mimicking the normal execution of a program.

\subsubsection{Normal Execution} 

To provide an intuitive insight into abstract interpretation, let us
consider how an execution of goal $g_{0}:append(L_1,L_2,L_3)$
transforms one program state ${\theta_0:\{L_1 \values [s(0)],L_2
\values [0]\}}$ which satisfies the condition in the above question
into another program state ${\theta_3:\{L_1 \values [s(0)], L_2
\values [0], L_3 \values [s(0),0]\}}$.  We deviate slightly from
Prolog-like logic programming systems. Firstly, substitutions (program
states) have been made explicit because the purpose of a program
analysis is to infer properties about substitutions. Secondly, when a
clause is selected to satisfy a goal with an input substitution, the
goal and the input substitution instead of the selected clause are
renamed. This is because we want to keep track of values of variables
occurring in the program rather than those of their renaming
instances.  Let $\Psi(Z)=Z'$ for any $Z\in\allvars$.

The first step performs a procedure entry.  ${g_0}$ and
 ${\theta_0}$ are renamed by into
 ${\Psi(g_0):append(L_1',L_2',L_3')}$ and ${\Psi(\theta_0):\{L_1'
 \values [s(0)],L_2' \values [0]\}}$.  Then C2 is
 selected and its head $append([H|L_1],L_2,[H|L_3])$ is unified
 with ${\Psi(g_0)}$ resulting in
 ${E_1:\{L_1' \values [H|L_1], L_2' \values L_2, L_3' \values
 [H|L_3]\}}$.  Then ${\Psi(\theta_0)}$ and ${E_1}$ are used to
 compute ${\theta_1:\{H \values s(0), L_1 \values [~], L_2 \values
 [0]\}}$ the input substitution of the body
 ${g_1:append(L_1,L_2,L_3)}$ of C2.  
In this way, the initial goal ${g_0}$ with
 its input substitution ${\theta_0}$ has been reduced into the
 goal ${g_1}$ and its input substitution ${\theta_1}$.

The execution of ${g_1}$, details of which have been omitted,
transforms ${\theta_1}$ into ${\theta_2:\{H \values s(0), L_1
\values [~], L_2 \values [0], L_3 \values [0]\}}$ the output
substitution of ${g_1}$.

The last step performs a procedure exit. The head of C2 and
${\theta_2}$ are renamed by $\Psi$, and then the renamed head
$append([H'|L_1'],L_2',[H'|L_3'])$ is unified with ${g_0}$ resulting
in  \mbox{${E_2: \{L_1\values [H'|L_1'],
L_2\values L_2', L_3\values [H'|L_3']\}}$}. Then $\Psi(\theta_2):\{H'
\values s(0), L_1' \values [~], L_2' \values [0], L_3' \values [0]\}$
and ${E_2}$ are used to update the input substitution ${\theta_0}$ of
${g_0}$ and this results in the output substitution ${\theta_3: \{L_1
\values [s(0)], L_2 \values [0], L_3 \values [s(0),0]\}}$ of ${g_0}$.

\subsubsection{Abstract Execution}
The above question is answered by an abstract execution of ${g_0}$
 which mimics the above normal execution of ${g_0}$.

The abstract execution differs from the normal execution in that
in place of a substitution is an abstract substitution that
describes groundness of the values that variables may take.  An
abstract substitution associates  a mode with a variable.  
$L \values \modeg$ means
that ${L}$ is a ground term and ${H \values\modeu}$
means that groundness of ${H}$ is unknown. Thus, the input abstract
substitution of $g_{0}$ is
$\mu_0:\{L_1\values\modeg,L_2\values\modeg\}$.

The first step of the abstract execution performs an abstract
procedure entry.  ${g_0}$ and ${\mu_0}$ are first renamed by
$\Psi$. Then C2 is selected and its head
$append([H|L_1],L_2,[H|L_3])$ is unified with ${\Psi(g_0)}$
resulting in ${E_1}$. Then
$\Psi(\mu_0):\{L_1'\values\modeg,L_2'\values\modeg\}$
and ${E_1}$ are used to compute the input abstract substitution
$\mu_1=\{H \values\modeg, L_1 \values \modeg, L_2 \values
\modeg\}$ of ${g_1}$ as follows.  $L_1'$ is ground 
by ${\Psi(\mu_0)}$ and $L_1'$ equals to $[H|L_1]$
by ${E_1}$. Therefore, ${[H|L_1]}$ is ground,
which implies that both $H$ and $L_1$ are ground. 
So, ${H \values\modeg}$ and $L_1 \values
\modeg$ are in ${\mu_1}$.  Similarly, $L_2 \values
\modeg$ is in ${\mu_1}$ since $L_2' \values L_2$ is in
${E_1}$ and $L_2' \values \modeg$ is in ${\Psi(\mu_0)}$.
In this way, the initial goal ${g_0}$ with its input abstract
substitution ${\mu_0}$ has been reduced into ${g_1}$ and its input
abstract substitution ${\mu_1}$. 

The abstract execution of ${g_1}$, details of which have been
omitted, transforms ${\mu_1}$ into ${\mu_2:\{H \values\modeg, L_1
\values \modeg, L_2 \values \modeg, L_3 \values
\modeg\}}$ the output abstract substitution of ${g_1}$.

 The last step of the abstract execution performs an abstract
procedure exit. The head of C2 and ${\mu_2}$ are renamed by $\Psi$,
and then the renamed head $append([H'|L_1'],L_2',[H'|L_3'])$ is
unified with ${g_0}$ resulting in ${E_2}$.  $\Psi(\mu_2):\{H \values
\modeg, L_1 \values \modeg, L_2 \values \modeg, L_3 \values \modeg\}$
and ${E_2}$ are then used to update the input abstract substitution
${\mu_0}$ of ${g_0}$ and this results in the output abstract
substitution $\mu_3:\{L_1\values \modeg, L_2 \values \modeg, L_3
\values \modeg\}$.  The modes assigned to ${L_1}$ and ${L_2}$ by
${\mu_3}$ are the same as those by ${\mu_0}$. By ${\Psi(\mu_2)}$, $H'$
and $L'$ are ground. Hence, $[H'|L']$ is ground.  By $E_2$, $L_3$
equals to $[H'|L']$ implying $L_3$ is ground.  So, $L_3\values \modeg$
is in ${\mu_3}$.

C1 may be applied at the first step since ${\mu_0}$ also describes
$\{L_1 \values [~],L_2 \values [0,s(0)]\}$.  This alternative abstract
computation would give the same output abstract substitution $\{L_1
\values \modeg, L_2 \values \modeg, L_3 \values \modeg\}$ of ${g_0}$.
Therefore, according to ${\mu_3}$, if ${L_1}$ and ${L_2}$ are ground
before ${append(L_1,L_2,L_3)}$ is executed, ${L_3}$ is ground after
${append(L_1,L_2,L_3)}$ is successfully executed.

The abstract execution closely resembles the normal
execution. They differ in that the abstract execution processes
abstract substitutions whilst the normal execution processes
substitutions and in that the abstract execution performs abstract
procedure entries and exits whilst the normal execution performs
procedure entries and exits.

\subsubsection{Abstract Interpretation} 
Performing program analysis by mimicking normal program execution
is called abstract interpretation.  The resemblance between normal
and abstract executions is 
common among different kinds of analysis.  An abstract interpretation
framework factors out common features of normal and abstract
executions and models normal and abstract executions by a number
of operators on a semantic domain, which leads to the following
formalisation of abstract interpretation.

Let $\ucdom$ be a set. An $n$-ary operator on $\ucdom$ is a
function from $\ucdom^{n}$ to $\ucdom$. An interpretation $\uci$
is a tuple $\langle
(\ucdom,\ucleq),(\ucop_1,\cdots,\ucop_k)\rangle $ where
$(\ucdom,\ucleq)$ is a complete lattice and
$\ucop_1,\cdots,\ucop_k$ are operators of fixed arities.
 
\begin{definition} \label{df:xmai}  
Let \[ \uci=\langle
(\ucdom,\ucleq),(\ucop_1,\cdots,\ucop_k)\rangle\]
\[\umai=\langle (\umdom,\umleq),(\umop_1,\cdots,\umop_k)\rangle\]
be two interpretations such that $\ucop_i$ and $\umop_i$ are of
same arity $n_i$ for each $1\leq i\leq k$, and $\umgamma$ be a
monotonic function from $\umdom$ to $\ucdom$. $\umai$ is called a
$\umgamma$-abstraction of $\uci$ if, for each $1\leq i\leq k$,
\begin{itemize} 
\item  
     for all $\Vector{\ucele}{{n_i}}\in \ucdom$ and
     $\Vector{\umele}{{n_i}}\in \umdom$, \[(\land_{1\leq k\leq{n_i}}
      \ucele_k \sqsubseteq
     \umgamma(\umele_k)) \implies
     \ucop_i(\Vector{\ucele}{{n_i}})\sqsubseteq \umgamma\circ
     \umop_i(\Vector{\umele}{{{n_i}}}) \]
 
\end{itemize}  
\end{definition}  
 
$\umai$ is called an abstract interpretation and $\uci$ a concrete
interpretation. An object in $\umai$, say $\umdom$, is called
abstract while an object in $\uci$, say $\ucdom$, is called
concrete. With respect to a given $\umgamma$, if
$\ucele\sqsubseteq\umgamma(\umele)$ then $\ucele$ is said to be
described by $\umele$ or $\umele$ is said to be a description of
$\ucele$. The condition in the above definition is read as that
$\umop_i$ is a $\umgamma$-abstraction of $\ucop_i$. Therefore,
$\umai$ is a $\umgamma$-abstraction of $\uci$ iff each operator
$\umop$ in $\umai$ is a $\umgamma$-abstraction of its
corresponding operator $\ucop$ in $\uci$.

\subsection{Abstract Interpretation Framework}

The abstract interpretation framework in~\cite{LU94} is based on a
collecting semantics of normal logic programs which associates
each textual program point with a set of substitutions. The set is
a superset of the set of substitutions that may be obtained when
the execution of the program reaches that program point. The
collecting semantics is defined in terms of two operators on
$(\wp(Sub),\subseteq)$. One operator is the set union $\cup$ - the
least upper bound operator on $(\wp(Sub),\subseteq)$ and the
other is $\cunify$ which is defined as follows. Let
$a_1,a_2\in\atom$ and $\Theta_1,\Theta_2\in\wp(Sub)$.
\[ 
\cunify(a_1,\Theta_1,a_2,\Theta_2) = 
\{\unify(a_1,\theta_1,a_2,\theta_2)\neq\fail~|~\theta_1\in\Theta_1
               \land\theta_2\in\Theta_2\} 
\] 
where
\begin{eqnarray*} 
\unify(a_1,\theta_1,a_2,\theta_2) & \definedas 
 & \begin{array}[t] {l} 
   \mgu((\Psi(\theta_1))(\Psi(a_1)),\theta_2(a_2))\circ\theta_2
  \end{array}  
\end{eqnarray*} 
which encompasses both procedure entries and procedure exits.  For
a procedure entry, $a_1$ is the calling goal, $\theta_1$ its input
substitution, $a_2$ the head of the selected clause and $\theta_2$
the empty substitution. For a procedure exit, $a_1$ is the head of
the selected clause, $\theta_1$ the output substitution of the
last goal of the body of the clause, $a_2$ the calling goal and
$\theta_2$ its input substitution.
 
The collecting semantics corresponds to the following
concrete interpretation.
\[ \langle (\csub,\cleq),(\club,\cunify)\rangle  
\]

Specialising the framework for a particular program analysis
consists in designing an abstract domain $(\fmsub,\fmleq)$, a
concretisation function $\fmgamma:\fmsub\mapsto\csub$, and two
\mbox{abstract} operators $\fmlub$ and $\fmunify$ such that
\mbox{$\langle (\fmsub,\fmleq),(\fmlub,\fmunify)\rangle $} is a
$\fmgamma$-abstraction of $\langle
(\csub,\cleq),(\club,\cunify)\rangle $. Once $(\fmsub,\fmleq)$ and
$\fmgamma$ are designed, it remains to design $\fmunify$ such that
$\fmunify$ is a $\fmgamma$-abstraction of $\cunify$ since $\fmlub$
is a $\fmgamma$-abstraction of $\club$.  $\fmunify$ is called an
abstract unification operator since its main work is to abstract
unification.  The abstract unification operator implements both
abstract procedure entries and abstract procedure exits as
$\unify$ encompasses both procedure entries and procedure
exits. We note that the \first{procedure-entry} and
\first{procedure-exit} operators in~\cite{Bruynooghe91}, the
\first{type substitution propagation} operator
in~\cite{Kanamori:93,KanamoriJLP93} and the corresponding
operators in~\cite{Muthukumar:JLP92} and~\cite{Nilsson:88} can be
thought of as variants of $\fmunify$.

\section{Polymorphic abstract interpretation} \label{sec:pai}
This section recalls the notion of polymorphic abstract interpretation
proposed in~\cite{LuJLP98}.

An analysis corresponds to an abstract interpretation.  For a monomorphic
analysis, data descriptions are monomorphic as they do not contain
parameters. For a polymorphic analysis, data descriptions contain
parameters.  Take the logic program in figure~\ref{fig:append} as an
example. By a monomorphic type analysis~\cite{HoriuchiK87,Lu95}, one would
infer \( {[L_{1}\in\tlist(\tnat),L_{2}\in\tlist(\tnat)]}
append(L_{1},L_{2},L_{3}){[L_{3}\in\tlist(\tnat)]} \) and \(
{[L_{1}\in\tlist(\tint),L_{2}\in\tlist(\tint)]}
append(L_{1},L_{2},L_{3}){[L_{3}\in\tlist(\tint)]} \).  It is desirable to
design a polymorphic type analysis~\cite{BarbutiG92,CodishD94,LuJLP98}
which would infer \({[L_{1}\in\tlist(\alpha),L_{2}\in\tlist(\alpha)]}
append(L_{1},L_{2},L_{3}){[L_{3}\in\tlist(\alpha)]} \).

The two statements inferred by the monomorphic type analysis are two
 instances of the statement inferred by the polymorphic type analysis.
 A polymorphic analysis is a representation of a possibly infinite
 number of monomorphic analyses.  The result of a polymorphic analysis
 subsumes the results of many monomorphic analyses in the sense that
 the results of the monomorphic analyses may be obtained as instances
 of the result of the polymorphic analysis. Let
 $\upai=<(\updom,\upleq),(\upop_1,\cdots,\upop_k)>$ be the abstract
 interpretation for a polymorphic analysis.  The elements in $\updom$
 by necessity contain parameters because the result of the polymorphic
 analysis contains parameters.  Since parameters may take as value any
 element from an underlying domain, it is necessary to take into
 account all possible assignments of values to parameters in order to
 formalise a polymorphic abstraction. In the sequel, an assignment
 will always mean an assignment of values to the paramenters that
 serve as placeholders for information to be provided after analysis.
 
\begin{definition}  \label{df:pai} Let $\upassign$ be the set of 
assignments, $\uci=<(\ucdom,\sqsubseteq),(\ucop_1,\cdots,\ucop_k)>$
and $\upai=<(\updom,\upleq),(\upop_1,\cdots,\upop_k)>$ be two
interpretations such that $\ucop_i$ and $\upop_i$ are of same arity
$n_{i}$ for each $1\leq i\leq k$, and
$\upgamma:\updom\times\upassign\mapsto \ucdom$ be monotonic in its
first argument. $\upai$ is called a polymorphic
$(\upgamma,\upassign)$-abstraction of $\uci$ if, for each $1\leq i\leq
k$,
\begin{itemize} \item 
for all $\kappa\in\upassign$, all $\Vector{\ucele}{{n_{i}}}\in \ucdom$ 
and all $\Vector{\upele}{{n_{i}}}\in \updom$,  
\[ (\land_{1\leq j\leq {n_i}} \ucele_j\sqsubseteq\upgamma(\upele_j,\kappa))
   \implies \ucop_i(\Vector{\ucele}{{n_{i}}})\sqsubseteq 
\upgamma ( \upop_i(\Vector{\upele}{{{n_{i}}}}),\kappa)
\]
\end{itemize} 
\end{definition}   
 
The above condition is read as that $\upop_i$ is a polymorphic
$(\upgamma,\upassign)$-abstraction of $\ucop_i$. Therefore, $\upai$ is
a polymorphic $(\upgamma,\upassign)$-abstraction of $\uci$ iff each
operator $\upop$ in $\upai$ is a polymorphic
$(\upgamma,\upassign)$-abstraction of its corresponding operator
$\ucop$ in $\uci$. With $\upassign$, $\upgamma$ and $\uci$ being understood,
$\upai$ is called a polymorphic abstract interpretation. The notion of
polymorphic abstract interpretation provides us better understanding
of polymorphic analyses and simplifies the design and the proof of
polymorphic analyses.

An alternative formulation of a polymorphic abstract interpretation is 
to define $\upgamma$ as a function of type 
$\updom\mapsto(\upassign\mapsto\ucdom)$. Thus, a polymorphic abstract
interpretation is a special class of abstract interpretation. 

\section {Monomorphic Groundness Analysis} 
\label{sec:mono} 
This section reformulates the groundness analysis presented
in~\cite{Sondergaard:86} that uses the abstract domain for groundness
proposed in~\cite{Mellish:85}. The reformulated groundness analysis will
be used section~\ref{sec:poly} to obtain the polymorphic groundness analysis.

\subsection{Abstract Domains}

In groundness analysis, we are interested in knowing which variables
in $\pvars$ will be definitely instantiated to ground terms and which
variables in $\pvars$ are not necessarily instantiated to ground
terms. We use $\modeg$ and $\modeu$ to represent these two
instantiation modes of a variable. Let
$\modes\definedas\{\modeg,\modeu\}$ and $\modeleq$ be defined as
$\modeg\modeleq\modeg$, $\modeg\modeleq\modeu$ and
$\modeu\modeleq\modeu$.  $\langle\modes,\modeleq\rangle$ is a complete
lattice with infimum $\modeg$ and supremum $\modeu$. Let $\modelub$
and $\modeglb$ be the least upper bound and the greatest lower bound
operators on $\langle\modes,\modeleq\rangle$ respectively.

The intuition of a mode in $\modes$ describing a set of terms is
captured by a  concretisation  function $\modecon:
\langle\modes,\modeleq\rangle\mapsto
\langle\wp(\sub),\subseteq\rangle$ defined in the following.

\[ \modecon(\mode) \definedas \left\{
   \begin{array}{ll} \term(\func,\emptyset), & if~\mode=\modeg\\
                     \term(\func,\allvars),  & if~\mode=\modeu
   \end{array}\right.
\]

A set of substitutions is described naturally by associating each
variable in $\pvars$ with a mode from $\modes$.  The
abstract domain for groundness analysis is thus
\(\langle\grddom,\grdleq\rangle\) where
\[\grddom\definedas \pvars\mapsto\modes
\]
and $\grdleq$ is the pointwise extension of $\modeleq$.
$(\grddom,\grdleq)$ is a complete lattice.  We use $\grdlub$
and $\grdglb$ to denote the least upper bound  and the
greatest lower bound operators on
\(\langle\grddom,\grdleq\rangle\) respectively.  The
set of substitutions described by a function from
$\pvars$ to $\modes$ is modelled by a concretisation
function $\grdcon$ from \mbox{$\langle\grddom,\grdleq\rangle$}
to $\langle\wp(\sub),\subseteq\rangle$ defined as follows.
\[
\grdcon (\mtheta)  \definedas 
  \{ \theta~|~\forall X\in\pvars. (\theta(X)\in\modecon(\mtheta(X))) \}
\]
$\grdcon$ is  a monotonic function from
$\langle\grddom,\grdleq\rangle$ to
$\langle\wp(Sub),\subseteq\rangle$.

Monomorphic abstract substitutions in $\grddom$ describes modes of
variables in $\pvars$. The abstract unification operator for the
monomorphic groundness analysis will also make use of modes of renamed
variables. Let $\pvars^{+}\definedas\pvars\cup\Psi(\pvars)$. We define
$\grddom^{+}\definedas\pvars^{+}\mapsto\modes$ and $\grdleq^{+}$ as
pointwise extension of $\modeleq$. $\grdcon^{+}$, $\grdlub^{+}$ and
$\grdglb^{+}$ are defined as counterparts of $\grdcon$, $\grdlub$ and
$\grdglb$ respectively.

\begin{lemma}
$\modecon(\modes), \grdcon(\grddom)$ and $\grdcon^{+}(\grddom^{+})$ are
Moore families.

\begin{proof} Straightforward. $\Box$
\end{proof}
\end{lemma}

\subsection{Abstract Unification Operator}

Algorithm~\ref{monoalgorithm} defines an abstract
unification operator for groundness analysis. Given
$\mtheta,\msigma\in \grddom$ and $A,B\in
ATOM(\func,\pred,\pvars)$, it computes $\munify(A,\mtheta, B,
\msigma)\in \grddom$ in five steps.  In step (1), $\Psi$ is
applied to $A$ and $\mtheta$ to obtain $\Psi(A)$ and $\Psi(\mtheta)$,
and $\Psi(\mtheta)$ and $\msigma$ are combined to obtain
$\mzeta=\Psi(\mtheta)\cup\msigma$ so that a substitution satisfying
$\mzeta$ satisfies both $\Psi(\mtheta)$ and $\msigma$. Note that
$\mzeta\in \grddom^{+}$. In step (2), $E_{0}=eq\circ \mgu(\Psi(A),B)$ is
computed. If $E_{0}=\fail$ then the algorithm returns
$\{X\mapsto\modeg~|~X\in\pvars\}$ - the infimum of
$\langle\grddom,\grdleq\rangle$.  Otherwise, the algorithm
continues. In step (3), $\meta=\mdown(E_{0},\mzeta)$ is computed so
that $\meta$ is satisfied by any $\zeta\circ mgu(E_{0}\zeta)$ for
any $\zeta$ satisfying $\mzeta$.  In step (4), the algorithm
computes $\mbeta = \mup(\meta,E_{0})$ from $\meta$ such that any
substitution satisfies $\mbeta$ if it satisfies $\meta$ and
unifies $E_{0}$. In step (5), the algorithm restricts $\mbeta$ to
$\pvars$ and returns the result.

\begin{algorithm} \label{monoalgorithm} 
  { Let $\mtheta,\msigma\in \grddom$,
$A,B\in\atom(\func,\pred,\pvars)$.
\begin{eqnarray*}
  \lefteqn{ \munify(A,\mtheta,B,\msigma)
    \definedas} \nonumber\\ & &
  \left\{\begin{array}{l}
        let~~~E_{0}=eq\circ\mgu(\Psi(A),B)~in\\
        if~~~~E_{0}\neq\fail\\
        then~
        \mup(E_{0},\mdown(E_{0},\Psi(\mtheta)\cup\msigma))\restrict\pvars\\
        else~~\{X\mapsto\modeg~|~X\in\pvars\}
     \end{array}\right. \\
  \lefteqn{\mdown(E,\mzeta) \definedas \meta}\\ 
         & where &  \meta(X)\definedas
           \left\{ \begin{array}{ll} 
                   \mzeta(X), & 
                        \multicolumn{1}{r}{\mbox{if $X\not\in range(E)$}}\\
                   \multicolumn{2}{l}{\mzeta(X)~\modeglb~
                   \modeglb_{(Y=t)\in{E}\land X\in\vars{t}}\mzeta(Y)},\\ 
			 \multicolumn{2}{r}{\mbox{otherwise.}}
                   \end{array} 
           \right. \\
  \lefteqn{\mup(E,\meta)  \definedas \mbeta}\\
         & where & \mbeta(X)\definedas
           \left\{ \begin{array}{ll} 
             \meta(X), & 
                \multicolumn{1}{r}{\mbox{if $X\not\in dom(E)$}}\\
             \multicolumn{2}{l}{\meta(X)~{\modeglb}~
                   {\modelub}_{Y\in\vars{E(X)}}\mzeta(Y)},\\ 
			 \multicolumn{2}{r}{\mbox{otherwise.}}
                   \end{array} 
           \right. \\
\end{eqnarray*}  
}
\end{algorithm} 

The following theorem states that the abstract unification
operator is a safe approximation of $\cunify$.

\begin{theorem} \label{monosafeness}  
For any $\mtheta,\msigma\in \grddom$ and any
  $A,B\in\atom(\func,\pred,\pvars)$.  \[
  \cunify(A,\grdcon(\mtheta), B,\grdcon(\msigma)) \subseteq \grdcon
  (\munify(A,\mtheta,B,\msigma)) \]

\begin{proof} 
  $\Psi(\mtheta)\cup\msigma\in\grddom^{+}$.  Let
  $\zeta\in\grdcon^{+} (\Psi(\mtheta)\cup\msigma)$ and
  $(Y\mapsto\modeg)\in\mdown(E_{0},\Psi(\mtheta)\cup\msigma)$. Then
  either $(Y\mapsto\modeg)\in \Psi(\mtheta)\cup\msigma$ or there
  is $X$ and $t$ such that
  $(X\mapsto\modeg)\in\Psi(\mtheta)\cup\msigma$, $(X=t)\in E_{0}$
  and $Y\in \vars{t}$. So, $Y(\zeta\circ mgu(E_{0}\zeta))$ is
  ground if $mgu(E_{0}\zeta)\neq\fail$. If every variable in a
  term is ground under a substitution then that term is ground
  under the same substitution. Therefore, if $(Z\mapsto\modeg)\in
  \mup(E_{0}, \mdown(E_{0},\Psi(\mtheta)\cup\msigma))$ then $Z$ is
  ground under $\zeta\circ mgu(E_{0}\zeta)$. This completes the
  proof of the theorem. $\Box$  \end{proof}
\end{theorem} 

\begin{example}{Let $\pvars=\{X,Y,Z\}$, 
   $A=g(X,f(Y,f(Z,Z)),Y)$, $B=g(f(X,Y),Z,X)$,
    $\mtheta=\{X\values\modeg,Y\values\modeu,Z\values\modeu\}$ and
    $\msigma=\{X\values\modeu,Y\values\modeu,Z\values\modeg\}$.
    $\munify(A,\mtheta,B,\msigma)$ is computed as follows.

    In step (1), $\Psi=\{X\values X_{0},Y\values Y_{0},Z\values
    Z_{0}\}$ is applied to $A$ and $\mtheta$.  \[
    \begin{array}{lll} \Psi(A) &=& g(X_0,f(Y_0,f(Z_0,Z_0)),Y_0)\\
    \Psi(\mtheta) & = &
    \{X_0\values\modeg,Y_0\values\modeu,Z_0\values\modeu\}
    \end{array} \] and 
    $\mzeta=\Psi(\mtheta)\cup\msigma=\{X_0\values\modeg,Y_0\values\modeu,Z_0\values\modeu,X\values\modeu,Y\values\modeu,Z\values\modeg\}$
    is computed.

        In step (2), $E_{0} =eq\circ mgu(\Psi(A),B) = \{X_{0}=f(Y_0,Y),
        Z=f(Y_0,f(Z_0,Z_0)), X=Y_0\}$ is computed. 

        In step (3), $\meta=
        \mdown(E_{0},\mzeta)=\{X_0\values\modeg,Y_0\values\modeg,
        Z_0\values\modeg,X\values\modeu,Y\values\modeg,Z\values\modeg\}$
        is computed.

        In step (4),
        $\mbeta=\mup(E_0,\meta)=\{X_0\values\modeg,Y_0\values\modeg,
        Z_0\values\modeg,X\values\modeg,Y\values\modeg,Z\values\modeg\}$ is
        computed.

        In step (5),
        $\mbeta\restrict\pvars=\{X\values\modeg,Y\values\modeg,
        Z\values\modeg\}$ is returned.

        So, $\munify(A,\mtheta,B,\msigma) =\{X\values\modeg,Y\values\modeg,
        Z\values\modeg\} $.
        $\Box$}
\end{example}

\section{Polymorphic Groundness Analysis} \label{sec:poly}
We now present the polymorphic groundness analysis. We first design
polymorphic domains corresponding to the monomorphic domains 
for the monomorphic
groundness analysis and then obtain the polymorphic groundness analysis
by replacing  two primitive monomorphic operators by their 
polymorphic counterparts. 

\subsection{Abstract Domains}
In a polymorphic groundness analysis, the input 
contains a number $\pars$ of mode parameters which may be filled
in with values from $\modes$. The set of assignments is
hence $\upassign=\pars\mapsto\modes$.

We first consider how to express mode information in presence of
mode parameters.  The polymorphic groundness analysis needs to
propagate mode parameters in a precise way. The abstract
unification operator for the monomorphic groundness analysis
computes the least upper bounds and the greatest lower bounds of
modes when it propagates mode information. This raises no
difficulty in the monomorphic groundness analysis as the two
operands of the least upper bound operator or the greatest lower
bound operator are modes from $\modes$. In the
polymorphic groundness analysis, their operands contains
parameters. This makes it clear that mode information can no
longer be represented by a single mode value or parameter in order
to propagate mode information in a precise manner. 

We use as a tentative polymorphic mode description a set of subsets of
 mode parameters. Thus, the set of tentative polymorphic mode
 descriptions is $\wp(\wp(\pars))$. The denotation of a set $\scal$ of
 subsets of mode parameters under a given assignment $\kappa$ is
 determined as follows. Let $\scal=\{S_1,S_2,\cdots,S_n\}$ and
 $S_i=\{\alpha_i^1,\alpha_i^2,\cdots,\alpha_i^{k_{i}}\}$. Then, for a
 particular assignment $\kappa$, $\scal$ is interpreted as
 $\modelub_{1\leq i\leq n}\modeglb_{1\leq j\leq k_{i}}
 \kappa(\alpha_i^j))$. 
 $\emptyset$ represents $\modeg$ under any assignment since
 $\modelub\emptyset=\modeg$ while 
$\{\emptyset\}$ represents $\modeu$ as $\modelub
 \modeglb\emptyset =\modeu$.  Let $\pmodepre$ be a relation on 
$\wp(\wp(\pars))$ defined as follows.
\[ \scal_1\pmodepre\scal_2\definedas \forall S_1\in\scal_1.\exists S_2\in\scal_2.(S_2\subseteq S_1)
\]

For instance, $\{\{\alpha_1,\alpha_2\},\{\alpha_1,\alpha_3\}\}\}
\pmodepre\{\{\alpha_1\}\}$ and\linebreak
$\{\{\alpha_1,\alpha_2\},\{\alpha_1,\alpha_3\}\}\}
\pmodepre\{\{\alpha_2\},\{\alpha_3\}\}$.  If $\scal_1\pmodepre\scal_2$
then, under any assignment, $\scal_1$ represents a mode that is
smaller than or equal to the mode represented by $\scal_2$ under the
same assignment. 
$\pmodepre$ is a pre-order.
It is reflexive and transitive but not antisymmetric. For instance,
$\{\{\alpha_1,\alpha_2\},\{\alpha_1\}\}\pmodepre\{\{\alpha_1\}\}$ and
$\{\{\alpha_1\}\}\pmodepre\{\{\alpha_1,\alpha_2\},\{\alpha_1\}\}$.
Define $\pmodeequiv$ as $\scal_1\pmodeequiv\scal_2\definedas
(\scal_1\pmodepre\scal_2)\land(\scal_2\pmodepre\scal_1)$.  Then
$\pmodeequiv$ is an equivalence relation on $\wp(\wp(\pars))$. 
The domain of mode descriptions is constructed as 
\begin{eqnarray*}
 \pmodes & \definedas & {\wp(\wp(\pars))}_{/\pmodeequiv}\\
 \pmodeleq & \definedas & {\pmodepre}_{/\pmodeequiv}
\end{eqnarray*}
$\langle\pmodes,\pmodeleq\rangle$ is a complete lattice with its
infimum being ${[\emptyset]}_{\pmodeequiv}$ and its supremum being
${[\{\emptyset\}]}_{\pmodeequiv}$. The least upper bound operator
$\pmodelub$ and the greatest lower bound operator $\pmodeglb$ on
$\langle\pmodes,\pmodeleq\rangle$ are given as follows. 
\begin{eqnarray*}
{[\scal_1]}_{\pmodeequiv} \pmodelub & {[\scal_2]}_{\pmodeequiv}
	\definedas & {[\scal_1\cup\scal_2]}_{\pmodeequiv}\\
{[\scal_1]}_{\pmodeequiv} \pmodeglb & {[\scal_2]}_{\pmodeequiv}
	\definedas & {[\{S_1\cup{S_2}~|~S_1\in\scal_1\land{S_2}\in
        \scal_2]\}}_{\pmodeequiv}
\end{eqnarray*}

The meaning of a polymorphic mode description is given by
$\pmodecon:\pmodes\times\upassign\mapsto\wp(\term(\func,\allvars))$
defined as follows.
\[ \pmodecon({[\scal]}_{\pmodeequiv},\kappa) \definedas 
\modecon(\modelub_{S\in\scal}\modeglb_{s\in S} \kappa(s))
\] 
$\pmodecon$ interprets $\scal$ as a disjunction of conjunctions of
mode parameters. For instance, if a variable $X$ is associated with a
mode description $\{\{\alpha_1,\alpha_2\},\{\alpha_1,\alpha_3\}\}$,
then its mode is $(\alpha_1
\modeglb\alpha_2)\modelub(\alpha_1\modeglb\alpha_3)$.  For simplicity,
a polymorphic mode description will be written as a set of subsets of
mode parameters, using a member of an equivalence class of
$\pmodeequiv$ to represent the equivalence class.

Polymorphic abstract substitutions are a function mapping a variable
in $\pvars$ to a mode description in $\pmodes$. Polymorphic abstract
substitutions are ordered pointwise. The domain of polymorphic
abstract substitutions is $\langle\pgrddom,\pgrdleq\rangle$ where
$\pgrddom\definedas\pvars\mapsto\pmodes$ and $\pgrdleq$ is the
pointwise extension of $\pmodeleq$.  $\langle\pgrddom,\pgrdleq\rangle$
is a complete lattice with its infimum being $\{X\mapsto \emptyset
~|~X\in\pvars\}$ and its supremum being $\{X\mapsto \{\emptyset\}
~|~X\in\pvars\}$. The least upper bound operator $\pgrdlub$ and the
greatest lower bound operator $\pgrdglb$ are pointwise extensions of
$\pmodelub$ and $\pmodeglb$ respectively. The meaning of a polymorphic
abstract substitution is given by
$\pgrdcon:\pgrddom\times\upassign\mapsto\wp(\sub)$ defined as follows.

\[
\pgrdcon (\ptheta,\kappa)  \definedas 
  \{ \theta~|~\forall X\in\pvars. (\theta(X)\in\pmodecon(\ptheta(X),\kappa)) \}
\]

We define $\pgrddom^{+}\definedas\pvars^{+}\mapsto\pmodes$ and
$\pgrdleq^{+}$ as pointwise extension of
$\pmodeleq$. $\pgrdcon^{+}$, $\pgrdlub^{+}$ and $\pgrdglb^{+}$ are
defined as counterparts of $\pgrdcon$, $\pgrdlub$ and $\pgrdglb$
respectively.

\begin{lemma}
$\pmodecon(\pmodes), \pgrdcon(\pgrddom)$ 
and $\pgrdcon^{+}(\pgrddom^{+})$ are Moore families.

\begin{proof} For any $\kappa\in\upassign$, 
we have $\pmodecon(  \{\emptyset\} ) = \sub$, implying
that $\pmodecon$ is co-strict. Let $\scal_1,\scal_2\in\pmodes$ and 
$\kappa\in\upassign$. Then either 
(a) $\pmodecon(\scal_1\pmodeglb\scal_2,\kappa)=\modecon(\modeg)$ 
or (b) $\pmodecon(\scal_1\pmodeglb\scal_2,\kappa)=\modecon(\modeu)$. 
In the case (a), we have that 
$\modeglb_{\alpha\in{S_1}}\kappa(\alpha)$ for any $S_1\in\scal_1$ and 
$\modeglb_{\beta\in{S_2}}\kappa(\beta)$ for any $S_2\in\scal_2$. This
implies that $\pmodecon(\scal_1,\kappa)=\modecon(\modeg)$ 
and $\pmodecon(\scal_2,\kappa)=\modecon(\modeg)$. Therefore,
$\pmodecon(\scal_1\pmodeglb\scal_2,\kappa)=\pmodecon(\scal_1,\kappa)\cap
 \pmodecon(\scal_2,\kappa)$ holds in the case (a). 
That $\pmodecon(\scal_1\pmodeglb\scal_2,\kappa)=\pmodecon(\scal_1,\kappa)\cap
 \pmodecon(\scal_2,\kappa)$ holds in the case (b) can be proven similarly. 
So, $\pmodecon(\pmodes)$ is a Moore family.

That $\pgrdcon(\pgrddom)$ and $\pgrdcon^{+}(\pgrddom^{+})$ are Moore families is an 
immediate consequence of that $\pmodecon(\pmodes)$ is a Moore family.
\end{proof}
\end{lemma}

\subsection{Abstract Unification Operator}

Algorithm~\ref{polyalgorithm} defines an abstract unification
operator for the polymorphic groundness analysis. It 
is obtained from that for the monomorphic groundness analysis by
replacing monomorphic descriptions with polymorphic descriptions,
$\modelub$ and $\modeglb$ by $\pmodelub$ and $\pmodeglb$ respectively, and 
renaming $\munify,\mdown$ and $\mup$ into $\punify,\pdown$ and $\pup$ 
respectively. 

\begin{algorithm} \label{polyalgorithm} 
  { Let $\ptheta,\psigma\in \pgrddom$,
$A,B\in\atom(\func,\pred,\pvars)$.
\begin{eqnarray*}
  \lefteqn{ \punify(A,\ptheta,B,\psigma)
    \definedas} \nonumber\\ & &
  \left\{\begin{array}{l}
        let~~~E_{0}=eq\circ\mgu(\Psi(A),B)~in\\
        if~~~~E_{0}\neq\fail\\
        then~
        \pup(E_{0},\pdown(E_{0},\Psi(\ptheta)\cup\psigma))
          {\restrict}\pvars\\
        else~~\{X\mapsto  \emptyset ~|~X\in\pvars\}
     \end{array}\right. \\
  \lefteqn{\pdown(E,\pzeta) \definedas \peta}\\ 
         & where &  \peta(X)\definedas
           \left\{ \begin{array}{ll} 
                   \pzeta(X), & 
                        \multicolumn{1}{r}{\mbox{if $X\not\in range(E)$}}\\
                   \multicolumn{2}{l}{\pzeta(X)~\pmodeglb~
                   \pmodeglb_{(Y=t)\in{E}\land X\in\vars{t}}\pzeta(Y)},\\ 
			 \multicolumn{2}{r}{\mbox{otherwise.}}
                   \end{array} 
           \right. \\
  \lefteqn{\pup(E,\peta)  \definedas \pbeta}\\
         & where & \pbeta(X)\definedas
           \left\{ \begin{array}{ll} 
             \peta(X), & 
                \multicolumn{1}{r}{\mbox{if $X\not\in dom(E)$}}\\
             \multicolumn{2}{l}{\peta(X)~{\pmodeglb}~
                   {\pmodelub}_{Y\in\vars{E(X)}}\pzeta(Y)},\\ 
			 \multicolumn{2}{r}{\mbox{otherwise.}}
                   \end{array} 
           \right. \\
\end{eqnarray*}  
}
\end{algorithm} 

\begin{example}{ Let 
   \begin{eqnarray*} 
     \pvars & = &\{X,Y,Z\}\\
     A &= & g(X,f(Y,f(Z,Z)),Y)\\
     B &= & g(f(X,Y),Z,X)\\
     \ptheta &=& \{X\values  \{\{\alpha_1,\alpha_2\}\} ,
	           Y\values  \{\{\alpha_1,\alpha_3\}\} ,
                   Z\values  \{\{\alpha_2,\alpha_3\}\} \}\\
     \psigma &=& \{X\values  \{\{\alpha_1\},\{\alpha_2\}\} ,
                   Y\values  \{\{\alpha_2,\alpha_3\}\} ,
                   Z\values  \{\emptyset\} \}
    \end{eqnarray*}
    $\punify(A,\ptheta,B,\psigma)$ is computed as follows.

    In step (1), $\Psi=\{X\values X_{0},Y\values Y_{0},Z\values
    Z_{0}\}$ is applied to $A$ and $\ptheta$.  \[
    \begin{array}{lll} \Psi(A) &=& g(X_0,f(Y_0,f(Z_0,Z_0)),Y_0)\\
    \Psi(\ptheta) & = &
   \{X_0\values  \{\{\alpha_1,\alpha_2\}\} ,
      Y_0\values  \{\{\alpha_1,\alpha_3\}\} ,
      Z_0\values  \{\{\alpha_2,\alpha_3\}\} \}
    \end{array} \] and
    $\pzeta=\Psi(\ptheta)\cup\psigma=
      \{X_0\values  \{\{\alpha_1,\alpha_2\}\} ,Y_0\values  \{\{\alpha_1,\alpha_3\}\} ,Z_0\values  \{\{\alpha_2,\alpha_3\}\} ,
X\values  \{\{\alpha_1\},\{\alpha_2\}\} ,Y\values  \{\{\alpha_2,\alpha_3\}\} ,Z\values  \{\emptyset\} \}$
    is computed.

        In step (2), $E_{0} = eq\circ\mgu(\Psi(A),B) = \{X_{0}=f(Y_0,Y),
        Z=f(Y_0,f(Z_0,Z_0)), X=Y_0\}$ is computed. 

        Step (3) computes
       \[\begin{array}{lcl} \peta &=& \pdown(E_{0},\pzeta)\\ 
                           &=&
        \left\{\begin{array}{l} 
        X_0\values  \{\{\alpha_1,\alpha_2\}\} ,
        Y_0\values  \{\{\alpha_1,\alpha_2,\alpha_3\}\} ,
        Z_0\values  \{\{\alpha_2,\alpha_3\}\} ,\\
        X\values  \{\{\alpha_1\},\{\alpha_2\}\} ,
        Y\values  \{\{\alpha_1,\alpha_2,\alpha_3\}\} ,
        Z\values  \{\emptyset\} \end{array}\right\}
         \end{array}\]

        Step (4) computes
        \[\begin{array}{lcl}
          \pbeta &=& \pup(E_0,\peta)\\
                 &=&
       \left\{\begin{array}{l} 
        X_0\values  \{\{\alpha_1,\alpha_2,\alpha_3\}\} ,
         Y_0\values  \{\{\alpha_1,\alpha_2,\alpha_3\}\} ,
        Z_0\values  \{\{\alpha_2,\alpha_3\}\} ,\\
        X\values  \{\{\alpha_1,\alpha_2,\alpha_3\}\} ,
        Y\values  \{\{\alpha_1,\alpha_2,\alpha_3\}\} ,
        Z\values  \{\{\alpha_2,\alpha_3\}  \end{array}\right\} 
        \end{array}\]

        In step (5), $\pbeta\restrict\pvars=\{X\values
        \{\{\alpha_1,\alpha_2,\alpha_3\}\} , Y\values
        \{\{\alpha_1,\alpha_2,\alpha_3\}\} , Z\values
        \{\{\alpha_2,\alpha_3\}\} \}$ is returned.  So,
        $\punify(A,\ptheta,B,\psigma) =$\linebreak $\{X\values
        \{\{\alpha_1,\alpha_2,\alpha_3\}\} , Y\values
        \{\{\alpha_1,\alpha_2,\alpha_3\}\} , Z\values
        \{\{\alpha_2,\alpha_3\}\} \}$.  $\Box$}
\end{example}

We now prove that for any assignment, the instantiated result of the
polymorphic groundness analysis is as precise as that of the
monomorphic groundness analysis corresponding to the assignment.

\begin{theorem} \label{polyprecision}
Let $  \scal_1 ,  \scal_1 \in
\pmodes$, $\mode_1,\mode_2\in\modes$, $\ptheta,\psigma\in\pgrddom$,
$\peta,\pzeta\in\pgrddom^{+}$,
$\mtheta,\msigma\in\grddom$,$\meta,\mzeta\in\grddom^{+}$,
$A,B\in\atom(\func,\pred,\pvars)$ and $E\in\wp(\eqs)$. 
For any $\kappa\in\upassign$,
\begin{itemize}
\item [(a)] if $\pmodecon(  \scal_1 ,\kappa)=\modecon(\mode_1)$ 
      and $\pmodecon(  \scal_2 ,\kappa)=\modecon(\mode_2)$ 
      then both $\pmodecon(  \scal_1 \pmodeglb (  \scal_2 
                 ,\kappa)=\modecon(\mode_1\modeglb\mode_2)$ 
      and $\pmodecon(  \scal_1 \pmodelub (  \scal_2 
                 ,\kappa)=\modecon(\mode_1\modelub\mode_2)$; 
\item [(b)] if $\pgrdcon^{+}(\pzeta,\kappa)=\grdcon^{+}(\mzeta)$ then 
      \[\pgrdcon^{+}(\pdown(E,\pzeta),\kappa)=\grdcon^{+}(\mdown(E,\mzeta))\]
\item [(c)] if $\pgrdcon^{+}(\peta,\kappa)=\grdcon^{+}(\meta)$ then 
      \[\pgrdcon^{+}(\pup(E,\peta),\kappa)=\grdcon^{+}(\mup(E,\meta))\] and
\item [(d)] if $\pgrdcon(\ptheta,\kappa)=\grdcon(\mtheta)$ and 
      $\pgrdcon(\psigma,\kappa)=\grdcon(\msigma)$ 
       then 
      \[\pgrdcon(\punify(A,\ptheta,B,\psigma),\kappa)=
       \grdcon(\munify(A,\mtheta,B,\msigma))\]
\end{itemize}
\begin{proof} (b) and (c) follow from (a) and they together imply (d). 
Thus, it remains to prove (a). 

Assume $\pmodecon( \scal_1 ,\kappa)=\modecon(\mode_1)$ and $\pmodecon(
\scal_2 ,\kappa)=\modecon(\mode_2)$. The proof of $\pmodecon( \scal_1
\pmodeglb ( \scal_2 , \kappa)=\modecon(\mode_1\modeglb\mode_2)$ is
done by considering four different combinations of values that
$\mode_1$ and $\mode_2$ can take. We only prove it for the case
$\mode_1=\modeg$ and $\mode2=\modeu$ as other cases are similar.
$\mode_1=\modeg$ implies
$\modeglb_{\alpha\in{S_1}}\kappa(\alpha)=\modeg$ for any
$S_1\in\scal_1$.  Thus $\modeglb_{\alpha\in{S_1\cup
S_2}}\kappa(\alpha)=\modeg$ for any $S_1\in\scal_1$ and
$S_2\in\scal_2$. So, $\pmodecon( \scal_1 \pmodeglb ( \scal_2 ,
,\kappa)=\modecon(\mode_1\modeglb\mode_2)$ as
$\mode_1\modeglb\mode_2=\modeg$. 

$\pmodecon( \scal_1 \pmodelub ( \scal_2
,\kappa)=\modecon(\mode_1\modelub\mode_2)$ can be proven similarly. $\Box$

\end{proof}
\end{theorem}

The following theorem states that the abstract unification operator
$\punify$ for the polymorphic groundness analysis is a safe
approximation of $\cunify$.

\begin{theorem} \label{polysafeness}  
For any $A,B\in\atom(\func,\pred,\pvars)$,
$\kappa\in\upassign$, and  $\ptheta,\psigma\in \pgrddom$,  \[
  \cunify(A,\pgrdcon(\ptheta,\kappa), B,\pgrdcon(\psigma,\kappa))
  \subseteq \pgrdcon (\punify(A,\ptheta,B,\psigma),\kappa) \]
	
\begin{proof} Let $\mtheta$ be such that $\grdcon(\mtheta)=
\pgrdcon(\ptheta,\kappa)$ and $\msigma$ be such that 
$\grdcon(\msigma)=\pgrdcon(\psigma,\kappa)$.  By theorem~\ref{monosafeness},
we have  
\[\cunify(A,\pgrdcon(\ptheta,\kappa), B,\pgrdcon(\psigma,\kappa))
  \subseteq \grdcon (\munify(A,\mtheta,B,\msigma))
\]
By theorem~\ref{polyprecision}, 
we have 
\[ \pgrdcon (\punify(A,\ptheta,B,\psigma),\kappa) =
   \grdcon (\munify(A,\mtheta,B,\msigma))
\]
Therefore, 
\[
  \cunify(A,\pgrdcon(\ptheta,\kappa), B,\pgrdcon(\psigma,\kappa))
  \subseteq \pgrdcon (\punify(A,\ptheta,B,\psigma),\kappa) 
\]
This completes the proof of the theorem. $\Box$
\end{proof}
\end{theorem}

\section{Implementation and Examples} \label{sec:imp}

We have implemented the abstract interpretation framework and the
polymorphic groundness analysis in SWI-Prolog. The abstract
interpretation framework is implemented using O'Keefe's least
fixed-point algorithm~\cite{OKeefe:87}. Both the abstract
interpretation framework and the polymorphic type analysis are
implemented as meta-interpreters using ground representations for
program variables and mode parameters.

\subsection{Examples}
The following examples present the results of the polymorphic groundness
analysis on some Prolog programs. The sets are represented by lists.
$V\mapsto T$ is written as $V/T$ in the results, $\alpha$ as ${\sl alpha}$,
and $\beta$ as ${\sl beta}$.

\begin{example} The following is the Prolog program from~\cite{SterlingS86} 
(p.\ 250) for looking up the value
for a given key in a dictionary represented as a binary tree and the result
of the polymorphic groundness analysis. 

{\small

\begin{verbatim}
:-     %[K/[[alpha]],D/[[beta]],V/[[gamma]]]
   lookup(K,D,V)
       %[K/[[alpha,beta]],D/[[beta]],V/[[beta,gamma]]].

lookup(K,dict(K,X,L,R),V) :-
        %[K/[[alpha,beta]],X/[[beta]],L/[[beta]],R/[[beta]],V/[[gamma]]],
    X = V,
        %[K/[[alpha,beta]],X/[[beta,gamma]],L/[[beta]],R/[[beta]],
        % V/[[beta,gamma]]].
lookup(K,dict(K1,X,L,R),V) :-
        %[K/[[alpha]],K1/[[beta]],X/[[beta]],L/[[beta]],R/[[beta]],
        % V/[[gamma]]],
    K < K1,
        %[K/[],K1/[],X/[[beta]],L/[[beta]],R/[[beta]],V/[[gamma]]],
    lookup(K,L,V),
        %[K/[],K1/[],X/[[beta]],L/[[beta]],R/[[beta]],V/[[beta,gamma]]].
lookup(K,dict(K1,X,L,R),V) :-
        %[K/[[alpha]],K1/[[beta]],X/[[beta]],L/[[beta]],R/[[beta]],
          V/[[gamma]]],
    K > K1,
        %[K/[],K1/[],X/[[beta]],L/[[beta]],R/[[beta]],V/[[gamma]]],
    lookup(K,R,V),
        %[K/[],K1/[],X/[[beta]],L/[[beta]],R/[[beta]],V/[[beta,gamma]]].
\end{verbatim}
}

The analysis is done for the goal $lookup(K,D,V)$ and the input
abstract substitution $\{K\mapsto\{\{\alpha\}\},
D\mapsto\{\{\beta\}\},V\mapsto\{\{\gamma\}\}\}$.  The input abstract
substitution states that before the goal is executed, the instantiation
mode of $K$ is $\alpha$, that of $D$ is $\beta$ and that of $V$ is
$\gamma$.

The analysis result indicates that, after the goal is executed, the
instantiation mode of $K$ is $\alpha\modeglb\beta$, that of $D$ remains
$\beta$ and that of $V$ becomes $\beta\modeglb\gamma$. This is captured
by the output abstract substitution. 

The result can be instantiated by one of eight assignments in
$\{\alpha,\beta,\gamma\}\mapsto\modes$. Under a given assignment, a
polymorphic mode description evaluates to a mode in $\{\modeg,\modeu\}$.
Let
$\kappa=\{\alpha\mapsto\modeg,\beta\mapsto\modeg,\gamma\mapsto\modeu\}$.
Under the assignment $\kappa$, the input abstraction substitution evaluates
to $\{K\mapsto\modeg,D\mapsto\modeg,V\mapsto\modeu\}$ and output
abstract substitution evaluates to
$\{K\mapsto\modeg,D\mapsto\modeg,V\mapsto\modeg\}$. This indicates
that if the goal $lookup(K,D,V)$ called with $K$ and $D$ being
ground then $V$ is ground after $lookup(K,D,V)$ is successfully
executed. 

The result also indicates that immediately before the built-in calls
$K<K1$ and $K>K1$ are executed, the instantiation modes of $K$
and $K1$ are respectively $\alpha$ and $\beta$. The Prolog requires that
both operands of the built-in predicates $"<"$ and $">"$ are ground before
their invocations. It is obvious that if $\alpha$ and $\beta$ are assigned
$\modeg$ then the requirement is satisfied and corresponding run-time
checks can be eliminated. Such inference of sufficient conditions on inputs
for safely removing run-time checks is enabled by the ability of 
 propagating parameters. Without propagation of
parameters, the inference would require a backward analysis.  Current
frameworks  for abstract interpretation of logic programs 
do not support backward analyses. $\Box$

\end{example}

\begin{example} \label{ex:permsort}
The following is the result of polymorphic groundness analysis of the
permutation sorting program from~\cite{SterlingS86} (p.\ 55).  The
analysis is done with the goal $sort(Xs,Ys)$ and the input abstract
substitution $\{Xs\mapsto\{\{\alpha\}\},Ys\mapsto\{\{\beta\}\}\}$. The
output abstract substitution obtained is
$\{Xs\mapsto\{\{\alpha\}\},Ys\mapsto\{\{\alpha,\beta \}\}\}$. This
indicates that the instantiation mode of $Ys$ after the successful
execution of $sort(Xs,Ys)$ is the greatest lower bound of the
instantiation modes of $Xs$ and $Ys$ prior to the execution of the
goal.

The abstract substitution associated with the program point
immediately before the built-in call $X=<Y$ associates both $X$
and $Y$ with the polymorphic mode description
$\{\{\alpha,\beta\}\}$.  For this mode description to evaluate to
$\modeg$, it is sufficient to assign $\modeg$ to either $\alpha$
or $\beta$.  

{\small
\begin{verbatim} 
:-     %[Xs/[[alpha]],Ys/[[beta]]]
   sort(Xs,Ys)
       %[Xs/[[alpha]],Ys/[[alpha,beta]]].

select(X,[X|Xs],Xs) :-
        %[X/[[alpha,beta]],Xs/[[alpha]]].
select(X,[Y|Ys],[Y|Zs]) :-
        %[X/[[beta]],Y/[[alpha]],Ys/[[alpha]],Zs/[[]]],
    select(X,Ys,Zs),
        %[X/[[alpha,beta]],Y/[[alpha]],Ys/[[alpha]],Zs/[[alpha]]].

ordered([]) :-
        %[].
ordered([X]) :-
        %[X/[[alpha,beta]]].
ordered([X,Y|Ys]) :-
        %[X/[[alpha,beta]],Y/[[alpha,beta]],Ys/[[alpha,beta]]],
    X =< Y,
        %[X/[],Y/[],Ys/[[alpha,beta]]],
    ordered([Y|Ys]),
        %[X/[],Y/[],Ys/[]].

permutation(Xs,[Z|Zs]) :-
        %[Xs/[[alpha]],Z/[[beta]],Zs/[[beta]],Ys/[[]]],
    select(Z,Xs,Ys),
        %[Xs/[[alpha]],Z/[[alpha,beta]],Zs/[[beta]],Ys/[[alpha]]],
    permutation(Ys,Zs),
        %[Xs/[[alpha]],Z/[[alpha,beta]],Zs/[[alpha,beta]],Ys/[[alpha]]].
permutation([],[]) :-
        %[].

sort(Xs,Ys) :-
        %[Xs/[[alpha]],Ys/[[beta]]],
    permutation(Xs,Ys),
        %[Xs/[[alpha]],Ys/[[alpha,beta]]],
    ordered(Ys),
        %[Xs/[[alpha]],Ys/[[alpha,beta]]].
\end{verbatim}
} $\Box$
\end{example}

\begin{example} 
The following is the pure \first{factorials} program
from~\cite{SterlingS86} (p.\ 39) together with the result of the
polymorphic groundness analysis. The program is analysed with the goal
$factorial(N,F)$ with input abstract substitution being
$\{N\mapsto\{\{\alpha\}\}, F\mapsto\{\{\beta\}\}\}$.  The result shows that
the goal $factorial(N,F)$ always succeeds with $N$ and $F$ being bound to
ground terms regardless of the instantiation modes of $N$ and $F$ before
the goal $factorial(N,F)$ is executed.

{\small
\begin{verbatim}
:-     %[N/[[alpha]],F/[[beta]]]
   factorial(N,F)
       %[N/[],F/[]].

natural_number(0) :-
        %[].
natural_number(s(X)) :-
        %[X/[]],
    natural_number(X),
        %[X/[]].

plus(0,X,X) :-
        %[X/[]],
    natural_number(X),
        %[X/[]].
plus(s(X),Y,s(Z)) :-
        %[X/[],Y/[],Z/[[]]],
    plus(X,Y,Z),
        %[X/[],Y/[],Z/[]].

times(0,X,0) :-
        %[X/[]].
times(s(X),Y,Z) :-
        %[X/[],Y/[],Z/[[]],XY/[[]]],
    times(X,Y,XY),
        %[X/[],Y/[],Z/[[]],XY/[]],
    plus(XY,Y,Z),
        %[X/[],Y/[],Z/[],XY/[]].

factorial(0,s(0)) :-
        %[].
factorial(s(N),F) :-
        %[N/[[alpha]],F/[[]],F1/[[]]],
    factorial(N,F1),
        %[N/[],F/[[]],F1/[]],
    times(s(N),F1,F),
        %[N/[],F/[],F1/[]].
\end{verbatim}
} $\Box$
\end{example}

\subsection{Performance}
The SWI-Prolog implementation of the polymorphic groundness analysis
has been tested with a set of benchmark programs that have been used
to evaluate program analyses of logic programs. The experiments are
done on a 5x86 IBM compatiable PC running Windows95.

The table in figure~\ref{fig:poly} illustrates the time performance of
the polymorphic groundness analysis. Every but the last row
corresponds to the result of the polymorphic groundness analysis of a
specific input.  The input consists of a program, a goal and an input
abstract substitution that specifies the modes of the variables in the
goal. The program and the goal are listed in the first and the third
column. The input abstract substitution is the most general for the
goal that associates each variable in the goal with a different mode
paramter. For instance, the abstract substitution for the first row is
$\{X\mapsto\{\{\alpha\}\},Y\mapsto\{\{\beta\}\}\}$.  The second column
lists the size of the program. The fourth column is the time in
seconds spent on the polymorphic groundness analysis of the input.
Last row gives the total size of the programs and the total time.

The table indicates that the prototype polymorphic groundness analyzer
spends an average of $0.0443$ seconds to process one line program.
This is an acceptable speed for many logic program programs.  Both the
abstract interpretation framework and the polymorphic groundness
analysis are implemented as meta-interpreters in a public domain
Prolog system.  Moreover, we use ground representations for program
variables and mode parameters. Using ground representations enables us
to make the prototype implementation in a short time and to avoid
possible difficulties that may arise due to improper use of meta-level
and object-level variables. However, it also prevents us from taking
advantages of built-in unification and forces us to code unification
in Prolog. We believe that both the use of meta-programming and that
of ground representations significantly slow the prototype. Therefore,
there is much space for improving the time performance of the
polymorphic groundness analysis through a better implementation.

\begin{figure}
{\small 
\begin{tabular}{|l|r|l|r|r|r|r|}\hline\hline
{\sf Program} & Lines & Goal & Poly & Mono & Ratio & Assignments\\
              &       &      & (sec) & (sec) & & \\\hline\hline
graph connectivity&32&connected(X,Y)&0.16&0.097&1.641&4\\\hline
merge&15&merge(Xs,Ys,Zs)&0.28&0.117&2.382&8\\\hline
buggy naive reverse&12&nrev(X,Y)&0.11&0.067&1.629&4\\\hline
buggy quick sort&33&qs(Li,Lo)&0.38&0.182&2.082&4\\\hline
improved quick sort&20&iqsort(Xs,Ys)&0.33&0.112&2.933&4\\\hline
tree sort&31&treesort(Xs,Ys)&0.44&0.122&3.591&4\\\hline
list difference&14&diff(X,Y,Z)&0.11&0.040&2.750&8\\\hline
list insertion&19&insert(X,Y,Z)&0.11&0.068&1.600&8\\\hline
quicksort with&20&quicksort(Xs,Ys)&0.28&0.110&2.545&4\\
difference list & & & & & &\\\hline
dictionary lookup&12&lookup(K,D,V)&0.17&0.062&2.720&8\\
in binary trees &&&&&&\\\hline
permutation sort&22&sort(Xs,Ys)&0.05&0.057&0.869&4\\\hline
heapify binary trees&25&heapify(Tree,Heap)&0.55&0.192&2.857&4\\\hline
exponentiation&23&exp(N,X,Y)&0.16&0.078&2.031&8\\
 by multiplication &&&&&&\\\hline
factorial&22&factorial(N,F)&0.06&0.067&0.888&4\\\hline
zebra&50&zebra(E,S,J,U,N,Z,W)&0.66&0.325&2.026&128\\\hline
tsp&148&tsp(N,V,M,S,C)&2.74&1.458&1.878&32\\\hline
chat&1040&chart\_parser&61.08&61.410&0.994&1\\\hline
neural&381&test(X,N)&6.86&5.510&1.243&4\\\hline
disj\_r&164&top(K)&1.82&1.505&1.209&2\\\hline
dnf&84&go&3.68&2.960&1.243&1\\\hline
rev&12&rev(Xs,Ys)&0.11&0.027&4.000&4\\\hline
tree order&34&v2t(X,Y,Z)&0.93&0.225&4.133&8\\\hline
serialize&45&go(S)&0.88&0.305&2.885&2\\\hline
cs\_r&314&pgenconfig(C)&12.53&9.910&1.264&2\\\hline
kalah&272&play(G,R)&4.01&3.212&1.248&4\\\hline
press&370&test\_press(X,Y)&28.94&15.325&1.888&4\\\hline
queens&29&queens(X,Y)&0.16&0.097&1.641&4\\\hline
read&437&read(X,Y)&35.09&26.710&1.313&4\\\hline
rotate&15&rotate(X,Y)&0.22&0.040&5.500&4\\\hline
ronp&101&puzzle(X)&2.20&1.015&2.167&2\\\hline
small&11&select\_cities(W,C1,C2,C3)&0.06&0.016&3.555&16\\\hline
peep&417&comppeepopt(Pi,O,Pr)&23.94&12.090&1.980&8\\\hline
gabriel&111&main(V1,V2)&2.14&1.112&1.923&4\\\hline
naughts-and-crosses&123&play(R)&1.15&0.880&1.306&2\\\hline
semi&184&go(N,T)&13.68&6.892&1.984&4\\\hline
&4642&&206.07&&2.168&9\\
\hline\hline\end{tabular} 
}
\caption{\label{fig:poly}Performance of Polymorphic Groundness Analysis}
\end{figure}

The same table compares the performance of the polymorphic groundness
analysis with that of the monomorphic groundness analysis presented
in~\cite{Sondergaard:86}. The monomorphic groundness analysis
in~\cite{Sondergaard:86} uses a subset of $\pvars$ as an abstract
substitution. The subset of $\pvars$ contains those variables that are
definitely ground under all concrete substitutions described by the
abstract substitution. This allows operators on abstract substitutions
to be optimised. The monomorphic groundness analysis is implemented in
the same way as the polymorphic groundness analysis.

The number of different assignments to the mode paramenters in the
input abstract substitution for the polymorphic groundness analysis is
two to the power of the number of the mode parameters. Each assignment
corresponds to a monomorphic groundness analysis that is performed and
measured.  The fifth column lists the average time in seconds spent on
these monomorphic groundness analyses. The sixth column list the ratio
of the fourth column by the fifth column. It is the ratio of the
performance of the polymorphic groundness analysis by that of the
monomorphic groundness analysis.  The seventh column lists the number
of the possible assignments for the mode parameters.

The table shows that the time the polymorphic groundness analysis
takes is from 0.869 to 5.500 times that the monomorphic groundness
analysis takes on the suit of programs. In average, the polymorphic
groundness analysis is 2.168 times slower.  This is due to the fact
that the polymorphic mode descriptions are more complex than the
monomorphic mode descriptions.  The abstract unification operator and
the least upper bound operator for the polymorphic groundness are more
costly than those for the monomorphic groundness analysis.

The result of the polymorphic groundness analysis is much more general
than that of the monomorphic groundness analysis. It can be
instantiated as many times as there are different assignments for the
mode parameters in the input abstract substitution. The average number
of different assignments is 9 which is 4.151 times the average
performance ratio.  This indicates that if all different monomorphic
groundness analyses corresponding to a polymorphic groundness analysis
are required, polymorphic groundness analysis is 4.151 times better.
Moreover, the seven rows with three mode paramenters have an average
performance ratio of 2.513 and 8 assignments. For these seven rows,
the polymorphic groundness analysis is 3.182 times better if all
different monomorphic groundness analyses corresponding to a
polymorphic groundness analysis are required.  Furthermore, the three
rows with four or more mode parameters have an average performance
ratio of 2.486 and an avarage of 58.666 different assignments.
Polymorphic groundness analysis is 23.598 times better for these three
rows if all different monomorphic groundness analyses are required.

\section{Conclusion} \label{sec:conc}
We have presented a new groundness analysis, called polymorphic
groundness analysis, that infers dependency of the groundness of the
variables at a program point on mode parameters that are input to the
groundness analysis and can be instantiated after analysis. The
polymorphic groundness analysis is obtain by simulating a monomorphic
groundness analysis.  Some experimental results with a prototype
implementation of the analysis are promising.  The polymorphic
groundness analysis is proved to be as precise as the monomorphic
groundness analysis.

The monomorphic groundness analysis that we consider in this paper is the
least powerful one. It uses a simple domain for groundness. As groundness
is useful both in compile-time program optimisations itself and in
improving precisions of other program analyses such as
sharing~\cite{Sondergaard:86,CortesiF91,MuthukumarH91,SundararajanC92,Jacobs:JLP92},
more powerful domains for groundness have been studied. These domains
consists of propositional formulae over program variables that act as
propositional variables. Dart uses the domain {\sf Def} of definite
propositional formulae to capture groundness dependency between
variables~\cite{Dart91}. For instance, the definite propositional formula
$x\leftrightarrow (y\land z)$ represents the groundness dependency that $x$
is bound to a ground term if and only if $y$ and $z$ are bound to ground
terms. {\sf Def} consists of propositional formulae whose models are closed
under set intersection~\cite{CortesiFW_JLP96}. Marriott and S{\o}ndergaard
use the domain ${\sf Pos}$ of positive propositional
formulae~\cite{Mar-Son:loplas93}. A positive propositional formula is true
whenever each propositional variable it contains is true. ${\sf Pos}$ is
strictly more powerful than ${\sf Def}$. It has been further studied 
in~\cite{CortesiFW_JLP96,Arm-Mar-Sch-Son:SAS94,Arm-Mar-Sch-Son:SCP98} and
has several
implementations~\cite{Charlier92,LeCharlier:1993:GAP,pvh:jlp95b,Codish:ILPS93,CD95:prop}.

The domain for the monomorphic groundness analysis in this paper
is isomorphic to a subdomain ${\sf Con}$ of ${\sf
Pos}$. ${\sf Con}$ consists of propositional formulae that are conjunctions
of propositional variables.

The polymorphic groundness analysis infers the dependencies of the
groundness of variables of interest at a program point on mode
parameters while a {\sf Pos}-based groundness analysis infers
groundness dependencies among variables of interest at a program
point. {\sf Pos}-based groundness analysis can also be used to infer
groundness dependencies between variables at a program point and
variables in the goal. But this requires a complex program
transformation that introduces a new predicate for each program point
of interest and may lead to solving large boolean equations involving
many propositional variables. Also, the less powerful domain {\sf Def}
cannot be used to infer groundness dependencies between variables at a
program point and variables in the goal because, unlike {\sf Pos},
{\sf Def} is not condensing~\cite{Jacobs:JLP92,Mar-Son:loplas93}.

Though the polymorphic groundness analysis is not intended for
inferring groundness dependencies among variables of interest at a
program point, it captures this kind of dependency indirectly. In
example~\ref{ex:permsort}, the output abstract substitution for the
goal $sort(Xs,Ys)$ is
$\{Xs\mapsto\{\{\alpha\}\},Ys\mapsto\{\{\alpha,\beta\}\}\}$. This
implies that whenever $Xs$ is bound to a ground term, $Ys$ is bound to
a ground term because assigning $\modeg$ to $\alpha$ will evaluate
$\{\{\alpha,\beta\}\}\}$ to $\modeg$ regardless of the mode assigned
to $\beta$. In general, if the abstract substitution at a program
point assign $\rcal_{j}$ to $Y_{j}$ for $1\leq{j}\leq{\l}$ and
$\scal_{i}$ to $X_{i}$ for $1\leq i\leq k$ and $\pmodelub_{1\leq j\leq
{\l}}\rcal_{j}\pmodeleq \pmodelub_{1\leq i\leq k}\scal_{k}$ then the
{\sf Pos} like proposition $\land_{1\leq i\leq k} X_{i}
\rightarrow\land_{1\leq j\leq\l} Y_{j}$ holds at the program point.
Thus the polymorphic groundness analysis can also infer the groundness
dependencies produced by a {\sf Pos}-based groundness
analysis. Moreover, it can be directly plugged into most abstract
interpretation frameworks for logic programs.  However, this kind of
groundness dependency in the result of the polymorphic groundness
analysis is not as explicit as in the result of a {\sf Pos}-based
analysis.

%\bibliography{referenc}

\end{document}